\documentclass[
    final            
  ]
  {aipproc}

\usepackage{amsmath, amsfonts, amssymb, natbib}
\layoutstyle{6x9}

\begin{document}

\title{The environmental dependence of neutral hydrogen content in spiral galaxies}

\classification{98.65.-r}
\keywords{Galaxy clusters}

\author{Jesse Miner}{
  address={Department of Physics and Astronomy, 
    University of North Carolina , Chapel Hill, NC, USA}
}
\author{Jim Rose}{
  address={Department of Physics and Astronomy, 
    University of North Carolina , Chapel Hill, NC, USA}
}
\author{Sheila Kannappan}{
  address={Department of Physics and Astronomy, 
    University of North Carolina , Chapel Hill, NC, USA}
}

\def\arcs{\char'175\ }
\def\arcsc{\char'175 }
\def\arcm{\char'023\ }
\def\arcmn{\char'023 }
\def\eg{e.g.,\ }
\def\ie{i.e.,\ }
\def\etal{et~al.\ }
\def\etalc{et~al.,\ }
\def\hub{\ifmmode H_\circ\else H$_\circ$\fi}
\def\kms{~km~s$^{-1}$\ }

\begin{abstract}
We present a study of the relationship between the deficiency of neutral 
hydrogen and the local three-dimensional number density of 
spiral galaxies in the Arecibo catalog \cite{springob05b} of global HI 
measurements. We find that the dependence on density of the HI content
is weak at low densities, but increases sharply at high densities where
interactions between galaxies and the intra-cluster
medium become important. This behavior is reminiscent of the 
morphology-density relation \cite{dressler80} in that the effect
manifests itself only at cluster-type densities, and indeed when we plot both
the HI deficiency-density and morphology-density relations, we see
that the densities at which they ``turn up'' are similar. This 
suggests that the 
physical mechanisms responsible for the increase in early types in 
clusters are also responsible for the decrease in HI content.

\end{abstract}

\maketitle

\section{Introduction}\label{intro}

Many observational studies (e.g. \cite{dressler80}) have shown 
that galaxy populations in clusters have very different properties than 
those in lower density environments. Of
note is the decrease in atomic hydrogen 
content of galaxies in clusters \cite{haynes84,solanes96}.
Recent studies by \cite{levy07} have shown that 
spirals in the Pegasus cluster, which is a relatively poor cluster, 
suffer from atomic hydrogen deficiency even though the densities are
relatively low. The question naturally arises how these spirals have
lost a large fraction of their gas while residing in an intermediate
density cluster. Therefore, we chose to investigate the behavior of HI content 
as a function of environment, and to specifically 
learn what, if any, dependence existed at intermediate
densities comparable to the Pegasus cluster. Thus, we aimed to explore
the characteristic density at which we see a
transition from ``normal'' field galaxies into gas-depleted cluster galaxies,
and to explore the behavior of the transition: is it a sudden increase or
a gradual trend that extends down to low densities (\ie that of loose
groups)?

\section{Galaxy Sample}\label{galsamp}

Our HI measurements were provided by the Arecibo catalog published by 
\cite{springob05b}. The catalog consists of global HI fluxes from over 8,000
spiral galaxies in the ``Arecibo sky.'' We calculate the HI deficiency 
parameter (DEF) using the technique of 
\cite{solanes96}. A galaxy's DEF value is the logarithmic difference
between the typical HI mass of a sample of field galaxies with the
same linear diameter and morphological type, and the observed HI mass
of the galaxy, and is thus a measure of the amount of atomic hydrogen
which has been removed from, or prevented from replenishing, the galaxy.

\section{Density Measurement}\label{dens}

We have chosen the three dimensional number density of galaxies to represent
the local environment, using the 
Updated Zwicky Catalog \cite{falco99}. With the sky positions and 
local-group--corrected redshifts of the 
objects in the UZC, and adopting a value of the Hubble 
constant of \hub = 75 km s$^{-1}$, we can assign each galaxy a 3D position in a
Cartesian coordinate system. To calculate the local density, we find the
mean distance to an object's six nearest neighbors, and use that distance to
define the radius of a sphere containing the ``local'' region. The number
of objects contained within the sphere is divided by the physical volume of
the sphere to obtain a number density in units of Mpc$^{-3}$. In order to
offset luminosity function (LF) completeness effects and peculiar velocity
contamination of redshifts, we have chosen a velocity range for our sample 
of 3,000 km s$^{-1}$ to 6,500 km s$^{-1}$, and have multiplied all 
of the densities by a LF correction factor. The factor is the ratio of
the observable number of galaxies at 3,000 km s$^{-1}$ to the observable
number of galaxies at the redshift of the galaxy in question, 
as measured by integrating the LF to the limiting magnitude of the 
sample.

\section{Results}

\begin{figure}[!h]
\includegraphics[scale=0.4]{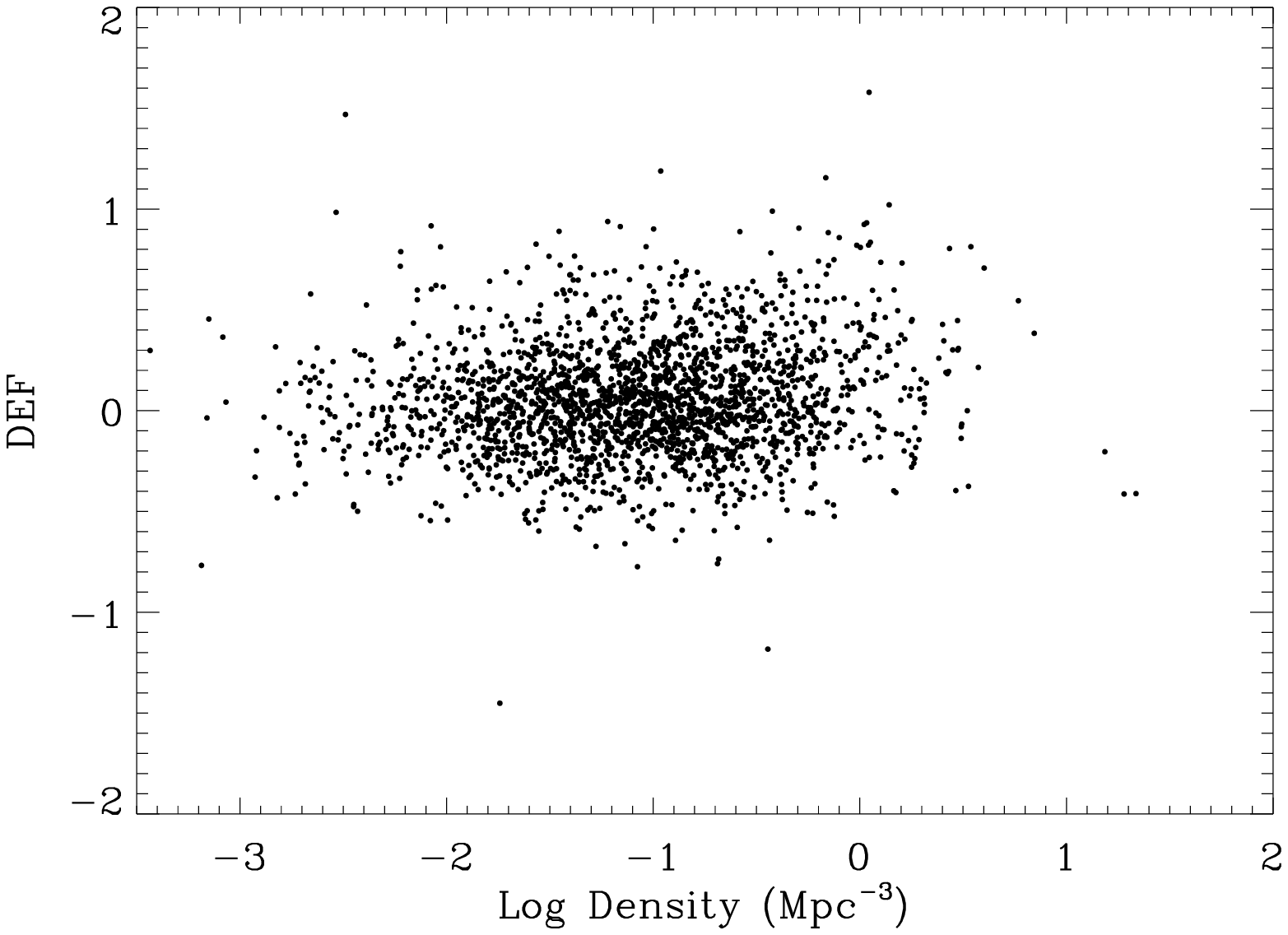}
\includegraphics[scale=0.4]{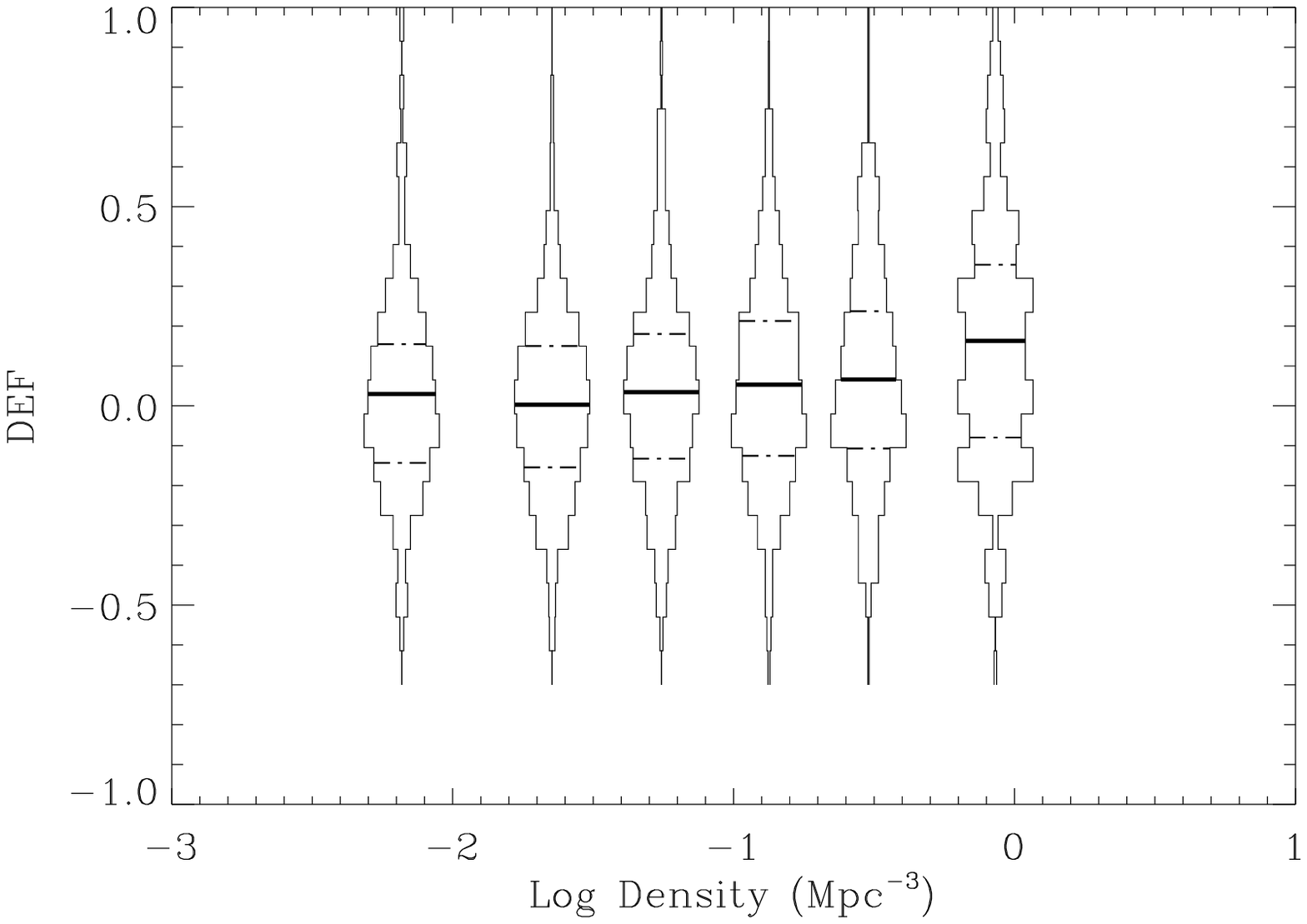}
\caption{\small Left: HI deficiency as a function of local density for 
our sample of 2,227 galaxies. Right: Points in top plot have been binned
in intervals of density. Thick lines are mean values, and dotted lines
are 25\% and 75\% quartiles.
\label{defdenspts}}
\end{figure}

We present a scatter plot of DEF vs density in Fig. \ref{defdenspts} for
our galaxy sample. It is difficult to make out any appreciable trend
by eye, so we also present the data binned in density intervals. 
It is clear from the figure that the mean DEF is relatively
constant, aside from some small scatter, at low and moderate densities, 
but increases suddenly in the highest density interval.

\begin{figure}[!h]
\includegraphics[scale=0.4]{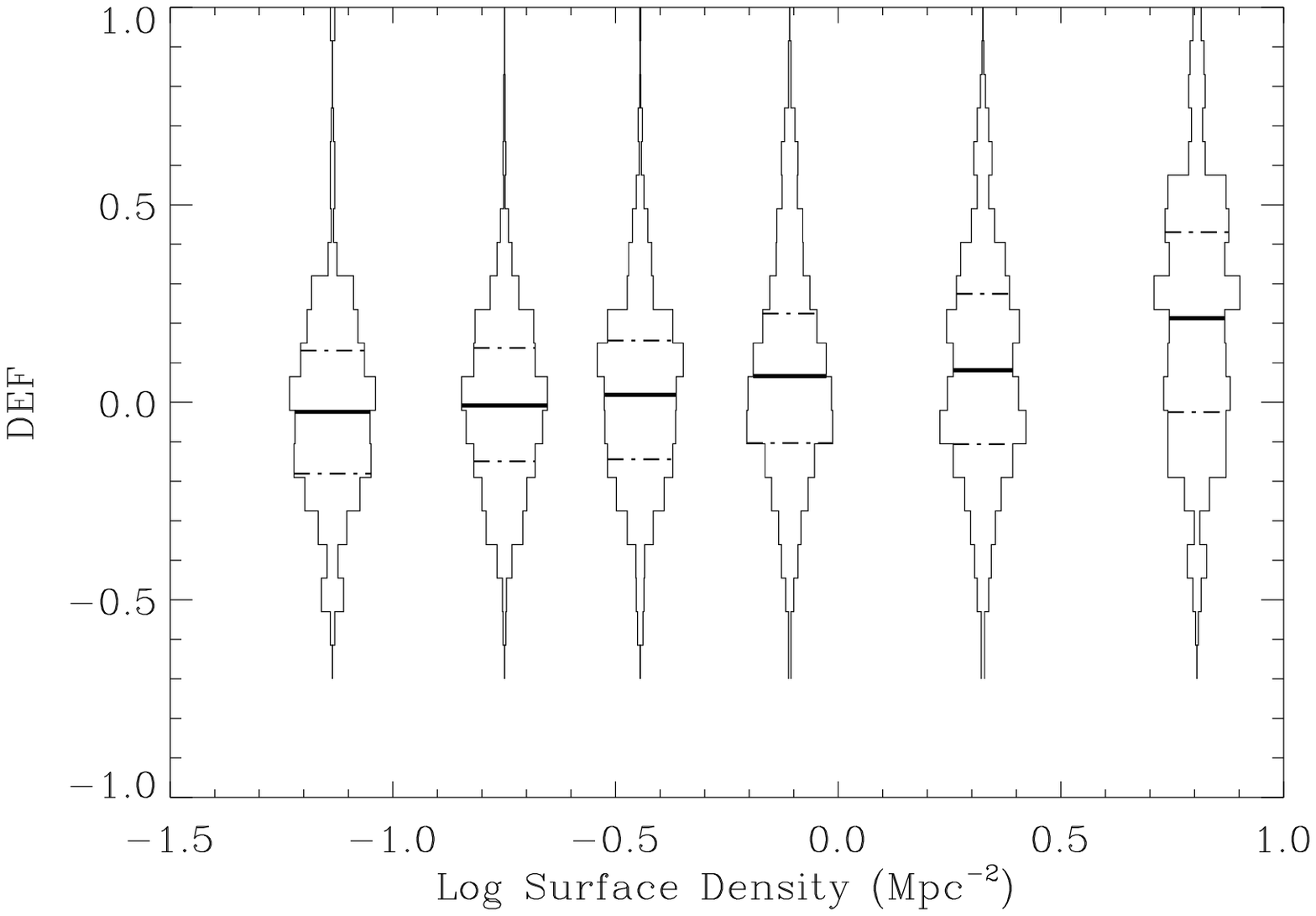}
\includegraphics[scale=0.4]{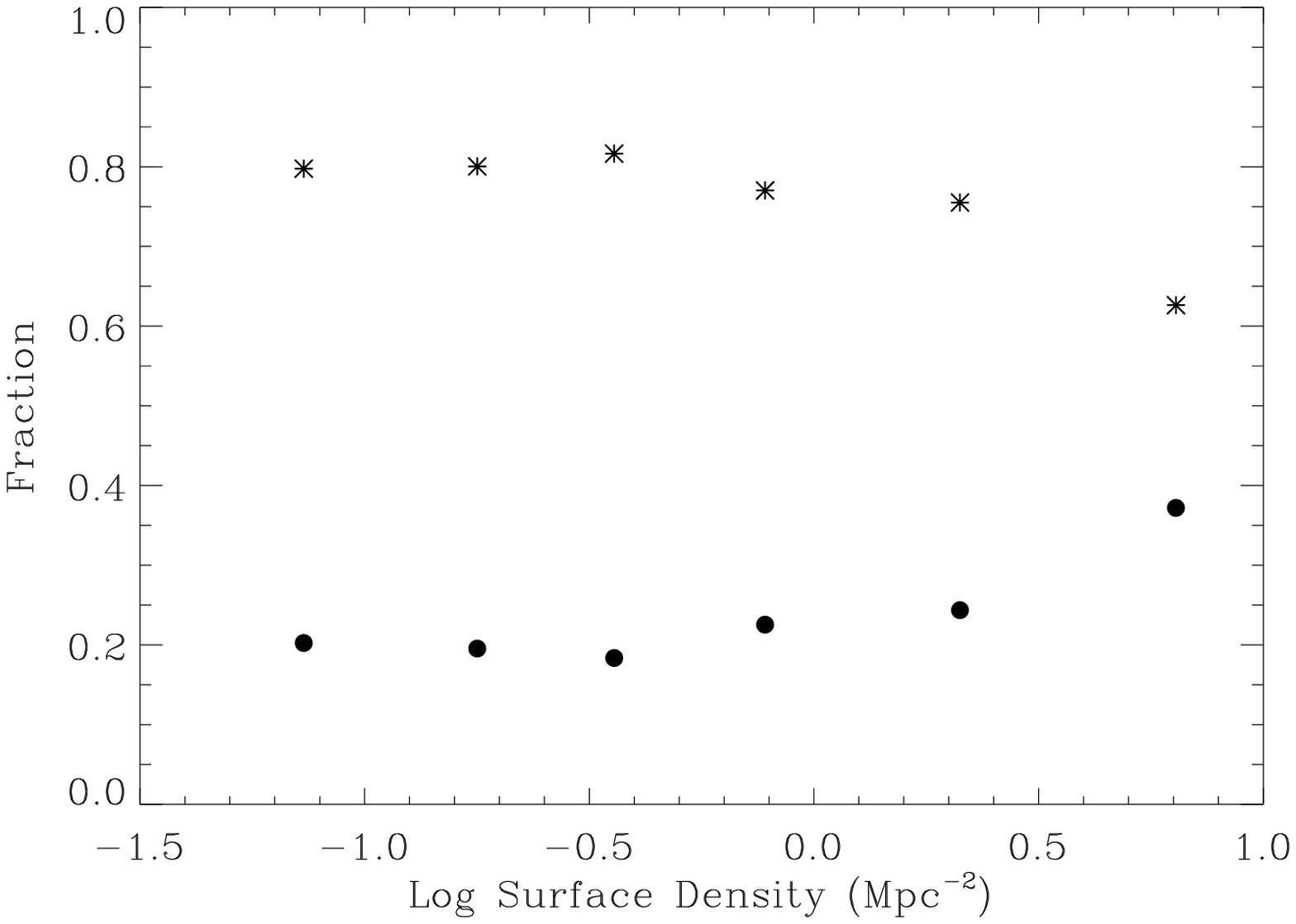}
\caption{\small Left: DEF-density relation using surface density. 
Right: Morphology-density relation using same sample of galaxies. 
Stars represent spirals, and circles represent E and S0 
galaxies.\label{morphdens}}
\end{figure}

The behavior of the deficiency with density is quite similar to the classical
morphology-density relationship (\cite{dressler80}), 
in that it is
relatively constant up to a certain density threshold and then increases
rapidly. We therefore reproduce the morphology-density
relation with our data, using similar techniques to those in the literature
(\cite{tanaka04}),
and compare it to the DEF-density trend. In Fig. \ref{morphdens}, we present
again the DEF-density relation, but with a surface density measurement 
calculated using the technique of \cite{tanaka04}. We also
plot the fractions of morphological types as a function of surface density
to show that the increase in HI deficiency corresponds to the increase
in early types as surface density increases. This similarity 
certainly points to a
common physical process which occurs in dense environments; obvious
possibilities include stripping by the intracluster medium and 
galaxy-galaxy interactions.

\section{Conclusions and Further Work}

Our investigation into the dependence on environment of the atomic hydrogen
content of galaxies has reinforced the idea that gas depletion 
mechanisms
play an important role in the observed decrease of galaxy gas as local
density increases. Specifically, we see that the same density regime
corresponds to both the increase in early type fraction and the 
decrease of HI mass in spiral galaxies. In the future, we plan on 
improving our
density estimates by including galaxy masses, and we 
will also examine the possible effect of a 
Malmquist-type bias due to distant, high deficiency
galaxies falling below the HI detection limit.

\bibliography{miner}

\end{document}